\documentclass[twocolumn,pre]{revtex4}

\usepackage{graphics}
\usepackage{amsmath}
\usepackage{amssymb}
\usepackage{amsfonts}
\usepackage{bm}

\begin{document}

\title{Totally asymmetric exclusion process with long-range hopping}

\author{J. Szavits-Nossan}
\email{juraj@ifs.hr}
\author{K. Uzelac}
\email{katarina@ifs.hr}

\affiliation{Institute of Physics, POBox 304, Bijeni\v{c}ka cesta 46, HR-10001 Zagreb, Croatia}

\begin{abstract}
Generalization of the one-dimensional totally asymmetric exclusion process (TASEP) with open boundary conditions
in which particles are allowed to jump $l$ sites ahead with the probability $p_l\sim 1/l^{\sigma+1}$ is studied 
by Monte Carlo simulations and the domain-wall approach. 
 For $\sigma>1$ the standard TASEP phase diagram is recovered, but the density profiles 
near the transition lines 
display new features when $1<\sigma<2$.
At the first-order transition line, the domain-wall is localized and phase separation is observed.
In the maximum-current phase the profile has an algebraic decay with a $\sigma$-dependent exponent.
Within the $\sigma \leq 1$ regime, where the transitions are found to be absent, analytical results 
in the continuum mean-field approximation are derived in the limit $\sigma=-1$. 
\end{abstract}


\date{\today}

\maketitle

\section{Introduction}

Asymmetric exclusion process (ASEP) represents one of the basic models in studying the rich behavior of systems held 
far from equilibrium \cite{revasep}. It was first introduced in \cite{MacDonaldGibbsPipkin68,MacDonaldGibbsPipkin69} as a lattice
model for protein synthesis and since then has been related to a wide range of other nonequilibrium
phenomena, like the surface growth \cite{MeakinRamanlalSanderBall86,KimKosterlitz89} belonging to the KPZ universality class
\cite{KardarParisiZhang86} or traffic flow \cite{NagelSchreckenberg92}.
In one dimension, subject to open boundary conditions, ASEP exhibits nonequilibrium phase transitions of the first 
and second order, which cannot occur in one-dimensional systems that are in thermal equilibrium 
and include only short-range interactions.
The finding of the exact analytical solution to this model \cite{DDM92,SchutzDomany93,DerridaEvans93} 
contributed to a better understanding of general mechanisms leading to phase transitions in systems out of 
equilibrium \cite{Privman-ed97}.
A number of different extensions of this model have shown that the ASEP phase 
diagram is quite robust to various modifications \cite{EvansBlythe02}.

The formulation of a large-deviation function \cite{DerridaLebowitzSpeer01,DerridaLebowitzSpeer02}, as a 
nonequilibrium equivalent of the free energy, pointed out the importance of the effective long-range 
interactions appearing in these processes.
In the present paper we include the long-range effect into this model in an explicit way, and examine the 
consequences for the phase diagram and critical behavior at the transition lines. 
We investigate the generalization of TASEP in which long-range interactions are introduced by letting 
particles overpass each other by jumping any distance $1\leq l\leq L$ with the probability $p_l\sim l^{-\sigma-1}$. 
Similar extensions have been considered in the high-speed Nagel-Schreckenberg traffic
model \cite{NagelSchreckenberg92,CheybaniKerteszSchreckenberg00} where cars are allowed to speed up 
if there is more than one empty site in front. However, these models do not allow cars to pass each other, which is 
justified for modeling single-lane traffic. 

An example of transport where particles are allowed to overpass each other may be found in the movement of the genome
regulatory proteins along the DNA. Namely, proteins bind to non-specific sites at DNA and search for a specific
target site either by one-dimensional diffusion along the DNA (sliding) or by continuous dissociation from and re-association to DNA. Once dissociated, proteins undergo three-dimensional diffusion and eventually rebind to another site. This process 
is either \textit{macroscopic} (fully dissociated protein rebinds to a random site on DNA)
or \textit{microscopic} (dissociated protein remains within the range of the electrostatic potential of DNA and rebinds to one
of the several nearby sites) \cite{BergWinterHippel}. In terms of TASEP, the first case corresponds to the creation and 
annihilation of particles in the bulk with constant rates \cite{ParmeggianiFranoschFrey03}. In the second case, 
dissociation and re-association events are correlated in space, which can be described in TASEP through the possibility
for particles to travel larger distances in a single jump.

The paper is organized as follows: in Section II we describe the generalization of ASEP to the long-range hopping, along
with the definition of the open boundary conditions adjusted to include the hopping with extended range. In Section III we present 
numerical results obtained by Monte Carlo simulations and compare them to the expected phase diagram obtained 
from the current-density relation (fundamental diagram). In Section IV we proceed to investigate density profiles. Using the domain-wall 
approach, we determine the criterion for the phase separation in terms of the parameter $\sigma$. In the maximum-current phase,
using the appropriate scaling of density profiles, we find that the density profiles decay algebraically with the exponent
that depends continuously on $\sigma$ for $1<\sigma<2$, while the usual exponent $\frac{1}{2}$ from ASEP is recovered for
$\sigma>2$. Conclusions are given in Section V.
 
\section{Long-range hopping with exclusion}

We consider the totally asymmetric exclusion process (TASEP) on a one-dimensional chain consisting of $L$ sites where
each site $i$ is either occupied by a particle ($\tau_{i}=1$) or empty ($\tau_{i}=0$). Instead of only the nearest-neighbor hopping,
we allow each particle to jump from the site $i$ to \textit{any} empty site $j>i$ on the right with the probability $p_l$
which decays with distance $l=|i-j|$ as 

\begin{equation}\label{p}
p_l=\frac{1}{\zeta_L(\sigma+1)}\frac{1}{l^{\sigma+1}},
\end{equation}

\noindent where the normalization $\zeta_L(\sigma+1)=\sum_{i=1}^{L}i^{-\sigma-1}$ is the partial sum of the Riemann zeta
function. In the limit $\sigma\rightarrow\infty$, the jump probability (\ref{p}) becomes $p_l=\delta_{l,1}$ and the model reduces
to the standard TASEP with the nearest-neighbor hopping. 

In order to define open boundary conditions we first introduce the particle reservoirs on the left and right 
end of the chain, with constant densities $\alpha$ and $1-\beta$, respectively, comprising $L$ sites each. Since each particle can travel up to $L$ sites in one jump,
we allow any particle from the original chain to jump to the right reservoir. For example, the particle at the site $L$ can
jump to any site in the right reservoir, whereas the particle at the first site can jump only to the site $L+1$
(first site of the right reservoir). Therefore, the total probability $\beta_i$ that the particle will leave the chain from
the site $i$ is the sum of probabilities of all possible jumps to the right reservoir from the site $i$, multiplied by the
probability $\beta$ that a site in the right reservoir is not occupied:

\begin{equation}\label{beta_i}
\beta_{i}=\frac{\beta}{\zeta_{L}(\sigma+1)}\sum_{j=L-i+1}^{L}\frac{1}{j^{\sigma+1}}.
\end{equation}

\noindent Similarly, only the particles from the left reservoir that are within a distance $L$ from the site $i$ in the
chain can jump to the site $i$. Since the probability to find the particle at any site in the left reservoir is $\alpha$, the total
probability $\alpha_i$ that the particle from the left reservoir will enter the chain at the site $i$ is given by

\begin{equation}\label{alpha_i}
\alpha_{i}=\frac{\alpha}{\zeta_{L}(\sigma+1)}\sum_{j=i}^{L}\frac{1}{j^{\sigma+1}}.
\end{equation}

\noindent Due to the particle-hole symmetry that is present in our model, the probabilities $\alpha_i$ and $\beta_i$ are not
independent. Namely, particles jumping to the right can be seen as holes jumping to the left, which gives the relation
$\alpha_i=\alpha\beta_{L-i+1}/\beta$.

Within the continuous-time dynamics, the time evolution of the system is described by the master equation:

\begin{equation}\label{master}
\frac{\partial P(C,t)}{\partial t}=\sum_{C'}{W(C'\rightarrow C)P(C',t)}-\sum_{C'}{W(C\rightarrow C')P(C,t)},
\end{equation}

\noindent where $P(C,t)$ is the probability of the system being in the state $C$
(given by the particular configuration of particles $\{\tau_i|i=1,...,L\}$) at time $t$ and $W(C\rightarrow C')\Delta t$
($\Delta t\rightarrow 0$) is
the probability of the system going from the state $C$ to the state $C'$ in the time interval $[t,t+\Delta t]$. Given any state
$C=\{\tau\}$, let $C^{i,j}$ be the configuration obtained from $C$, with $\tau_{i}$ and $\tau_{j}$ exchanging places, and
$C^{i}$ the configuration obtained from $C$ with $\tau_{i}\rightarrow 1-\tau_{i}$. The system then evolves in time in two
ways: either by exchanging particles and holes with the reservoirs ($C\rightarrow C^{i}$) or by exchanging the particle
at the site $i$ with the hole at the site $j>i$ ($C\rightarrow C^{ij}$). The impact of each of these two contributions
depends on the interaction parameter $\sigma$. To each update mechanism we can assign a characteristic length defined as the
first moment of the corresponding jump probability distribution.

For the jumps \textit{within} the chain, we get the average distance $\lambda_L$ as

\begin{equation}\label{lambda}
\lambda_{L}(\sigma)=\sum_{l=1}^{L}lp_{l}=\frac{\zeta_{L}(\sigma)}{\zeta_{L}(\sigma+1)}.
\end{equation}

Similarly, the average distances $\gamma_-$ and $\gamma_+$ from the boundaries at which the particles are injected and removed
from the chain, respectively, are given by 

\begin{equation}
\gamma_-=\alpha\gamma_L(\sigma),\qquad \gamma_+=\beta\gamma_L(\sigma),
\end{equation}

\noindent where $\gamma_L(\sigma)$ is defined as

\begin{equation}\label{gamma}
\gamma_L(\sigma)=\frac{1}{\alpha}\sum_{l=1}^{L}l\alpha_l=\frac{1}{\beta}\sum_{l=1}^{L}l\beta_l=
\frac{\lambda_L(\sigma)}{2}\left[1+\frac{\zeta_L(\sigma-1)}{\zeta_L(\sigma)}\right].
\end{equation}

\noindent By comparing these two lengths in the limit $L\rightarrow\infty$, two different regions may be distinguished
with respect to the parameter $\sigma$: for $\sigma>2$, both $\lambda=\lim_{L\rightarrow\infty}\lambda_L$ and
$\gamma=\lim_{L\rightarrow\infty}\gamma_L$ are finite, whereas for
$1<\sigma<2$, $\lambda$ is finite, but $\gamma$ diverges. In the first case the particles, on average,
enter and leave the chain within some finite distance $\gamma$ from the boundaries. In the
second case, where $\gamma$ is infinite, particles are, on average, created and annihilated in the bulk and the model
is in the regime similar to the TASEP with the bulk reservoir \cite{ParmeggianiFranoschFrey03}. The results presented below
confirm this simple estimate.

From the master equation (\ref{master}), the time change of the average particle density $\langle\tau_i(t)\rangle$
at the site $i$ is given by the discrete lattice continuity equation,

\begin{equation}\label{dis_cont_eq}
\frac{d}{dt}\langle\tau_{i}(t)\rangle=J_{i}-J_{i+1},
\end{equation}

\noindent where $J_{i}$ is the total current of particles, both those passing over the site $i$ and those leaving it
\begin{equation}\label{totalcurrent}
J_{i}=\sum_{k=i+1}^{L}\alpha_{k}(1-\langle\tau_{k}\rangle)+\sum_{k=1}^{i}\sum_{l=i+1}^{L}p_{l-k}
\langle\tau_{k}(1-\tau_{l})\rangle+\sum_{k=1}^{i}\beta_{k}\langle\tau_{k}\rangle.
\end{equation}

\noindent In the stationary state, the current $J_{i}$ is site-independent, and we can simply write the total current
$J_{i}\equiv j_{L}(\alpha,\beta,\sigma)$ as a function of parameters $\alpha$, $\beta$, $\sigma$, and $L$. 

In order to make a comparison with TASEP, we first examine the dependence of the total current $j_{L}$ on 
parameters $\alpha$ and $\beta$, and possible discontinuities in its derivatives leading to nonequilibrium 
phase transitions.

\section{Phase diagram}

From the way the boundary conditions were constructed, we already know that the special choice of $\alpha$ and $\beta$
such that $\alpha=1-\beta$ is equivalent to the case where the boundary conditions are periodic. Since the translational
invariance is then restored, all configurations $C$ are equally probable \cite{MeakinRamanlalSanderBall86}. The
stationary density profile is flat, i.e. $\langle\tau_i\rangle=\alpha$, which upon inserting into the expression for the current
(\ref{totalcurrent}) gives

\begin{equation}
j_L(\alpha,\beta,\sigma)=\lambda_L(\sigma)\alpha(1-\alpha),\qquad \alpha+\beta=1.
\end{equation}

\begin{figure}[!hb]
\includegraphics{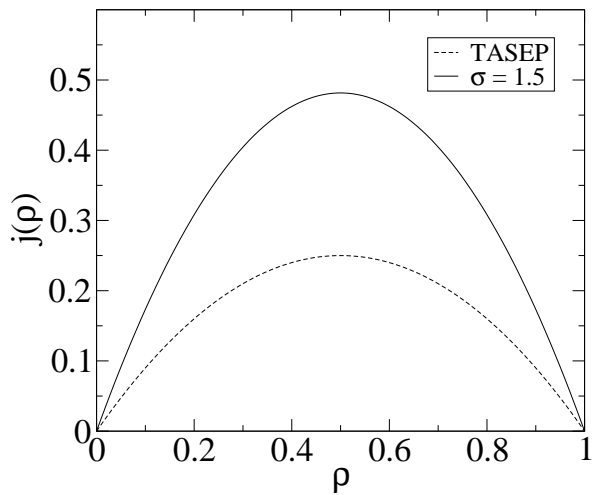}
\caption{Fundamental diagram of TASEP with long-range hopping with $\sigma=1.5$ (solid line)
and TASEP with the nearest-neighbor hopping (dashed line).}
\label{fig1}
\end{figure}

\noindent In the case of periodic boundary conditions ($\tau_i=\tau_{i+L}$), the number of particles $N$ is conserved, and
the average density profile $\langle\tau_i\rangle$ is constant and equal to $N/L$. This gives the current-density relation
(Fig. \ref{fig1})

\begin{equation}\label{funddiag}
j(\rho)=\lambda_{L}\rho(1-\rho).
\end{equation}\\

%
%

\noindent It has been argued \cite{Krug91, PopkovSchutz99}, that the knowledge of the
current-density relation $j(\rho)$ is sufficient to determine the phase diagram of a driven diffusive system coupled to
the left and right reservoirs with constant densities $\rho_{-}=\alpha$ and $\rho_{+}=1-\beta$, respectively. Explicitly, it was suggested
\cite{PopkovSchutz99} that the stationary current of the system with open boundaries, in case of a positive drift
($j\geq 0$), is always selected according to the extremal principle

\begin{equation}\label{extprinc}
j=\left\{\begin{array}{ll}
max\{j(\rho), \rho_{+}\leq\rho\leq\rho_{-}\} & \textrm{$\rho_{-}\geq\rho_{+}$} \\ \\
min\{j(\rho), \rho_{-}\leq\rho\leq\rho_{+}\} & \textrm{$\rho_{-}<\rho_{+}$.} 
\end{array}\right..
\end{equation}

\noindent In the present model, the current-density relation (\ref{funddiag}) has several important properties:
(i) except for the factor $\lambda_{L}(\sigma)$, which accounts for the average length of the jumps, it is the same as in 
TASEP; (ii) in the infinite system the factor $\lambda$, and therefore the current as well, 
are finite only for $\sigma>1$; (iii) $j(\rho)$ is symmetric to the particle-hole exchange. From the last property it follows that the
phase diagram is symmetric to the mutual exchange of the parameters $\alpha$ and $\beta$. If the extremal principle
applies to our model, the first property means that the structure of the phase diagram is the same as in TASEP. The second
property then ensures that the current is still well defined for an infinite system.

In the present model, the current has only one maximum at $\rho^{*}=1/2$, like in TASEP. Consequently, the expected phase diagram has
three phases: a maximum-current phase in the domain $\rho_{-}>\rho^{*}$, $\rho_{+}<\rho^{*}$, with the bulk density
$\rho=\rho^{*}$, a low-density phase in the domain $\rho_{-}<\rho^{*}$, with the bulk density $\rho=\rho_{-}$,
and a high-density phase in the domain $1-\rho_{+}>\rho^{*}$, with the bulk density $\rho=1-\rho_{+}$. 

The agreement of these predictions with the Monte Carlo simulations for different parameters $\alpha$, $\beta$,
and $\sigma$ is presented in the following section.

\subsection{Numerical results}

Monte Carlo (MC) simulations were performed with the random sequential
update. In other words, in each discrete time step ($L$ such steps making one Monte Carlo Step per site (MCS/site)), the site $i$ is randomly chosen
from the chain of $L$ sites. If the site $i$ is empty, a particle jumps from the left reservoir to the site $i$ with the
probability $\alpha_{i}$. If the site $i$ is occupied, an integer $m\in\{1,\dots,L\}$ is drawn from the probability
distribution $p_m$ and two cases are possible: in the first case, if the site $i+m\leq L$ is empty, the particle at the site
$i$ jumps to the site $i+m$, whereas in the second case, if $i+m>L$, the particle leaves the chain with the probability $\beta$. 
In a typical run of the simulation we calculated the time averages of density profiles and the particle current
$j_{L}$, omitting the results from the first $t_0$ MCS/site, where $t_0$ is the time required for the system to reach
the (quasi) stationary state. In exploring the phase diagram, a modest size $L=200$ was sufficient, and we
used $t\sim t_0\sim 10^6$ MCS/site. 

We found the same phase diagram as in TASEP only for $\sigma>1$, which is in agreement with the fact that $\lambda$
diverges for $\sigma\leq 1$. Since these two regions are essentially different, we treat them separately.\\

\begin{figure}[hb]
\centering\includegraphics{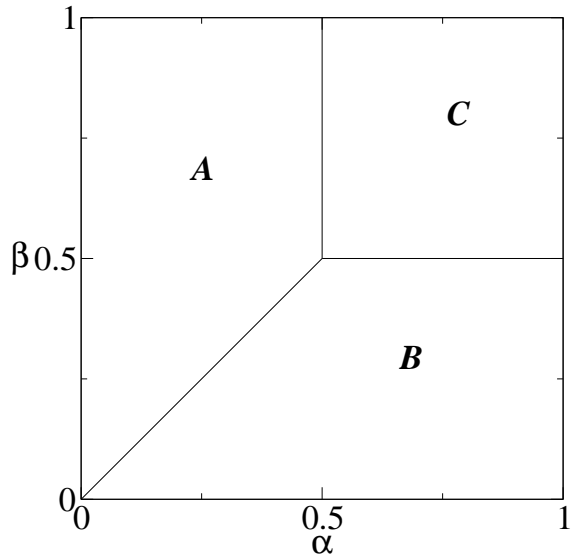}
\caption{Phase diagram for $\sigma>1$ from the results of Monte Carlo simulations consisting of low-density (A), high-density
 (B) and maximum-current (C) phases.}
\label{fig2}
\end{figure}

\begin{figure}
\includegraphics{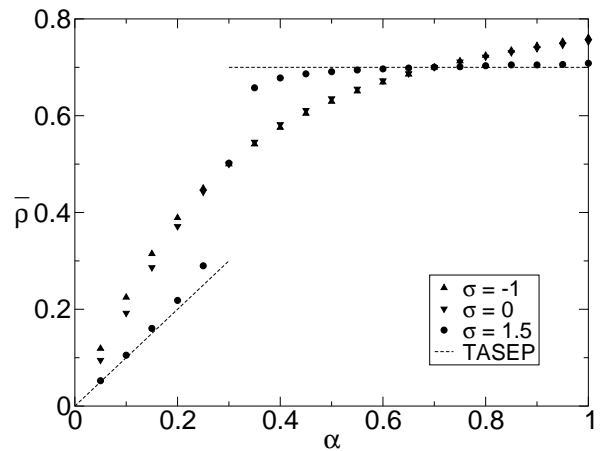}
\caption{Dependence of the bulk density $\overline{\rho}$ on $\alpha$ for
$\beta=0.3$ and various $\sigma$ ($L=200$, $t=2\cdot10^{6}$ MCS/site) compared to the bulk density in TASEP (dashed lines).}
\label{fig3}
\end{figure}

\begin{figure}
\includegraphics{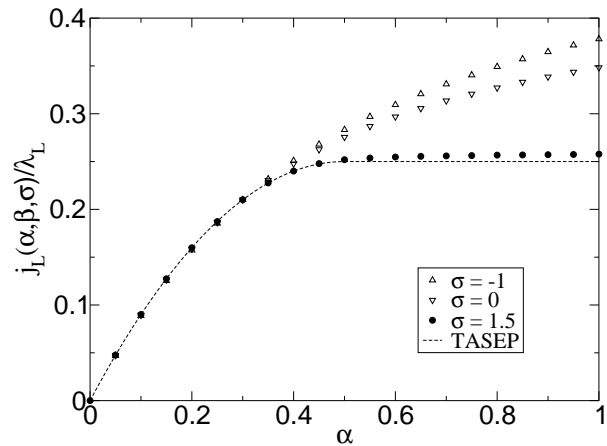}
\caption{Dependence of current $j_L(\alpha,\beta,\sigma)/\lambda_L$ on $\alpha$ for $\beta=0.7$ and various
$\sigma$ ($L=200$, $t=2\cdot10^{6}$ MCS/site) compared to the current in TASEP (dashed line).}
\label{fig4}
\end{figure}

\noindent \textbf{The case $\sigma>1$.} For $\sigma>1$ the results of the MC simulations, depending on parameters 
$\alpha$ and $\beta$, reveal three phases with the boundaries (Fig. \ref{fig2}) and bulk densities of each phase the same
as in TASEP \cite{SchutzDomany93,DerridaEvans93}. In order to check the order of the phase transitions across the boundaries, we consider the bulk density $\overline{\rho}=\langle N\rangle/L$ and the current $j_L$ as functions of $\alpha$ for  
two characteristic constant values of $\beta$ (Fig. \ref{fig3}, Fig. \ref{fig4}). Since the phase diagram is symmetric
to the exchange $\alpha\leftrightarrow\beta$, this is sufficient to cover the phase diagram in the whole region of the
parameters $0\leq\alpha,\beta\leq 1$.

%
%

In Fig. \ref{fig3} we observe that for $\sigma>1$ the bulk density has a discontinuity at the coexistence line
$\alpha=\beta$ between the low-density and high-density phases, which is characteristic of a first-order transition.
In Fig. \ref{fig4} the continuous transition to the maximum-current phase is presented. Although the
finite-size effects in the maximum-current phase are visible, this diagram is in agreement with the 
extremal principle (\ref{extprinc}) for $\rho^{*}=1/2$ and symmetric to the mutual exchange of parameters
$\alpha$ and $\beta$, as concluded above from symmetry considerations.\\

%
%

%
%

\noindent \textbf{The case $\sigma\leq 1$.} In this domain, the current diverges in the limit $L\rightarrow\infty$ and
it turns out that this completely destroys the phase diagram. The results from Monte Carlo simulations show the absence
of the first- and second-order phase transitions, as shown in Fig. \ref{fig3} (for varying $\alpha$ at constant 
$\beta=0.3$) and Fig. \ref{fig4} (for varying $\alpha$ at constant $\beta=0.7$).

In the limiting case $\sigma=-1$, a simple analytical solution of the stationary 
Eq. (\ref{dis_cont_eq}) may be derived in the continuum mean-field approximation 
for $\alpha=\beta$ (see Appendix A); i.e., by assuming 
$\langle\tau_i\tau_j\rangle=\langle\tau_i\rangle\langle\tau_j\rangle$ and setting $x=i/L$ in the limit
$L\rightarrow\infty$. In particular, the current can be written in the closed form as

\begin{equation}\label{jot_div}
j/L=\frac{\alpha(2\alpha+3)}{2}\left(1-\frac{2\alpha}{\sqrt{1-2\alpha}}
Arsh\sqrt{\frac{1-2\alpha}{2\alpha(2\alpha+3)}}\right).
\end{equation}

\noindent It should be pointed out that all derivatives of (\ref{jot_div}) are smooth functions of $0<\alpha=\beta\leq 1$. 

\section{Density profiles}

Monte Carlo simulations show that in the region $\sigma>1$, different phases have the same bulk density as in standard
TASEP. On the other hand, at the phase boundaries the spatial dependence of density profiles differs from the one
in TASEP and in general depends on $\sigma$. The exact solution of the present model is unknown and the 
mean-field approximation does not simplify it since it gives the complicated recursion equation for the average density
$\langle\tau_i\rangle$ that includes all the other densities $\langle\tau_j\rangle$, $j\neq i$. 
The only case we managed to solve in closed form is the limiting case $\sigma=-1$ in the continuum limit
(see Appendix A for $\alpha=\beta$), which is out of our main
scope due to the absence of phase transitions in the region $\sigma\leq 1$. 

In this section, we show how the domain-wall approach, previously used in TASEP \cite{Kolomeisky98}, can be
applied to the present model for the special choice of parameters $\alpha=\beta<1/2$ as long as $\sigma>1$ and 
treated analytically in order to compare it with the results from the simulations. 

In the maximum-current phase, to which this approach does not apply, the spatial 
dependence of the density profile is investigated by applying finite-size scaling
to the results obtained from the MC simulations.
 
\subsection{Line $\alpha=\beta$}

The domain-wall approach \cite{Kolomeisky98} in TASEP is a way to describe the collective motion of particles with 
a step-like shock profile performing the homogeneous random walk on a one-dimensional lattice consisting of $L+1$ sites. The
approach is based on the fact that the stationary continuity equation (\ref{dis_cont_eq}) in the
hydrodynamic limit, which is precisely the inviscid Burgers equation \cite{LebowitzPresuttiSpohn88}, has a multi-valued
solution at some point $x_s$. This can be described by the formation of a discontinuity (\textit{shock}) in a density
profile moving with the velocity $v_{s}$ \cite{LighthillWhitham55},

\begin{equation}\label{v_shock}
v_{s}=\frac{j[\rho(x_{s}^{+})]-j[\rho(x_{s}^{-})]}{\rho(x_{s}^{+})-\rho(x_{s}^{-})}.
\end{equation}

\noindent Depending on the sign of this velocity, the shock in TASEP is localized either at the left boundary
(high-density phase) or at the right boundary (low-density phase). For $\alpha=\beta$ the velocity $v_s$ vanishes 
and since the shock has equal probability to be at any site, the resulting density profile is linear.

In order to apply this approach to the present model, one has to assume a particular shock profile. In general, 
this can be done by
investigating the continuum mean-field approximation of the continuity equation (\ref{dis_cont_eq}) and constructing
the shock profile from the two solutions, each matching one boundary condition. Instead, for $\sigma>1$, we approximated
the instantaneous shock profile $\rho_s(x)$ with a step function, like it has been done in TASEP:

\begin{equation}\label{shock_profile}
\rho_{s}(x)=\left\{\begin{array}{ll}
\alpha & \textrm{$x<x_s$} \\ \\
1-\beta & \textrm{$x>x_s$.} 
\end{array}\right..
\end{equation}

\noindent This is justified by the fact that the phase diagram, determined by the bulk density in each phase, is the same
as in TASEP. However, the feature in which the present case differs from TASEP is that the hopping rates of the random walker
are site-dependent, because the excess current of all particles entering and leaving the 
low-density or the high-density domain depends on the position of the shock. This feature was already 
encountered in TASEP with the ``Langmuir kinetics'' \cite{EvansJuhaszSanten03}, in which particles
are created and annihilated in the bulk with constant rates $\Omega_C$ and $\Omega_D$, respectively.  

The hopping rate $r_i$ from the site $i$ to the site $i+1$ is proportional to the total current of particles entering
and leaving the high-density domain $\rho_{+}=1-\beta$: 

\begin{equation}\label{p_i}
r_{i}=\frac{1}{1-\alpha-\beta}\left[\sum_{j=i}^{L}\beta_{j}(1-\beta)-\sum_{j=i}^{L}\alpha_{j}\beta\right],\quad i=1,...,L.
\end{equation}

\noindent Similarly, the hopping rate $l_i$ from the site $i$ to the site $i-1$ is proportional to the total current
of particles entering and leaving the low-density domain $\rho_{-}=\alpha$:

\begin{equation}\label{q_i}
l_{i}=\frac{1}{1-\alpha-\beta}\left[\sum_{j=1}^{i-1}\alpha_{j}(1-\alpha)-\sum_{j=1}^{i-1}\beta_{j}\alpha\right],
\quad i=2,...,L+1.
\end{equation}

\noindent The shock is reflected at the boundaries, i.e. $l_1=0$ and $r_{L+1}=0$. 

\noindent If we compare the bias $v_i=r_i-l_i$ of the random walker in the present model with the one from TASEP
($v_i=\beta-\alpha$), we see that, as long as $\alpha \neq \beta$,  the qualitative picture from TASEP remains the same: 
for $\alpha<\beta$ we have $v_i>0$, $\forall i$ and the random walker is localized at the right end (low-density phase); 
for $\alpha>\beta$ we have $v_i<0$, $\forall i$ and the random walker is localized at the left end (high-density phase). 
However, on the coexistence line $\alpha=\beta$ the bias is not zero, except at the specific site 
$i=L/2+1$ ($x\approx 1/2$ for $L\gg 1$), where both hopping rates $r_i$ and $l_i$ are the same. 

The stationary probability $P_{i}$ to find the random walker at the site $i$ is 
given by the set of equations \cite{SantenAppert02}

\begin{equation}\label{equP_i}
\left\{\begin{array}{ll}
r_{i-1}P_{i-1}+l_{i+1}P_{i+1}-(r_i+l_i)P_{i}=0, & \textrm{$i=2,...,L$} \\ \\
l_{2}P_{2}-r_1P_{1}=0 \\ \\
r_{L}P_{L}+-l_{L+1}P_{L+1}=0,
\end{array}\right.
\end{equation}

\noindent where the last two equations account for the reflecting boundary conditions. The solution to the above
equations (\ref{equP_i}) can be expressed in the closed form as

\begin{equation}\label{P_i}
P_{i}=\frac{1}{Z_L}\prod_{j=1}^{i-1}{\frac{r_j}{l_{j+1}}}\equiv\frac{1}{Z_L}e^{-V_i},
\end{equation}

\noindent where $Z_L$ is the normalization. 
In the last expression we have rewritten $P_i$ in another form using the potential $V_i$, 

\begin{equation}\label{V_i}
V_i=-\sum_{j=1}^{i-1}ln\frac{r_j}{l_{j+1}},\qquad L+1\geq i>1.
\end{equation}

\noindent Once the probability $P_{i}$ is known, the density profile $\langle\tau_i\rangle^{DW}$ in the domain-wall 
approximation is simply given by \cite{SantenAppert02} 

\begin{equation}\label{rho_dw}
\langle\tau_i\rangle^{DW}=\left(\sum_{j=i+1}^{L+1}P_j\right)\alpha+\left(\sum_{j=1}^{i}P_j\right)(1-\beta).
\end{equation}

\begin{figure}
\includegraphics{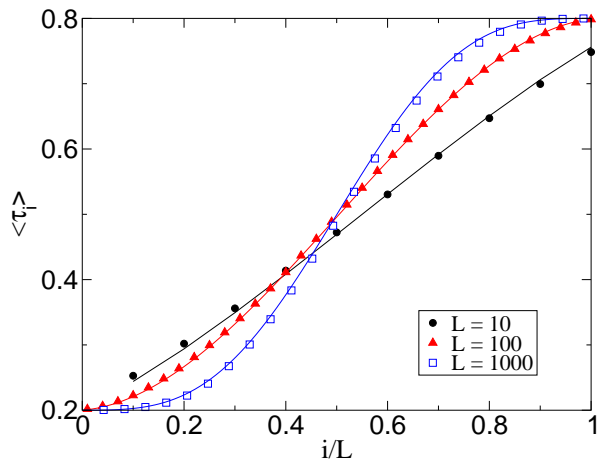}
\caption{(Color online) Comparison of the density profiles ($\alpha=\beta=0.2$, $\sigma=1.5$) for various 
chain sizes $L=10$, $100$,
and $1000$ obtained from MC simulations (symbols) and from the domain wall approach (solid lines).}
\label{fig5}
\end{figure}

\noindent The expression (\ref{rho_dw}) can be evaluated numerically and compared with the profiles obtained 
by MC simulations. 
In Fig. \ref{fig5} this was done for $\sigma=1.5$ and various system sizes ($L=10$, $100$, and $1000$). 
The agreement between the MC data and Eq. (\ref{rho_dw}) is excellent and justifies our initial assumption, 
even for very small system sizes ($L=10$). 
(A slight discrepancy  between Eq. (\ref{rho_dw}) and the actual profile obtained by MC simulations was 
observed only for $\sigma$ close to 1, and it could be reduced by increasing the size of the system.)

%
%

The nonlinear shape of the profile becomes more pronounced as $L$ increases.
For example, in Figure \ref{fig5}, the difference between the density profile for $\sigma=1.5$ and the linear profile
is more evident for larger system sizes, suggesting that in the thermodynamic limit the shock 
tends to localize in the middle of the chain.

\begin{figure}
\includegraphics{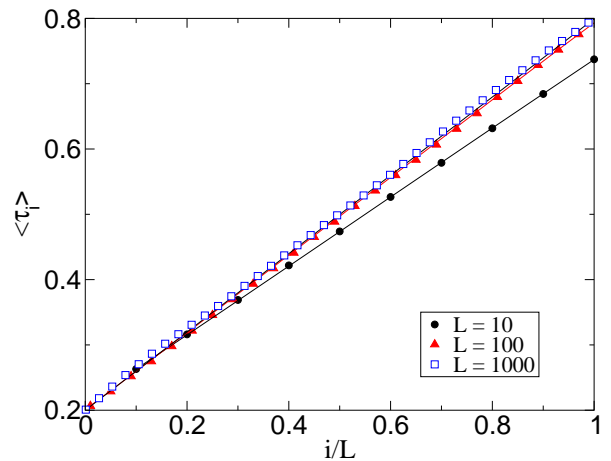}
\caption{(Color online) Comparison of the density profiles ($\alpha=\beta=0.2$, $\sigma=3.0$) for various chain sizes
$L=10$, $100$, and $1000$ obtained from simulations (symbols) and from the domain-wall approach (solid lines).}
\label{fig6}
\end{figure}

As we increase the parameter $\sigma$, the deviation from the linear profile is less visible. 
In the limit $\sigma\rightarrow\infty$, which is exactly TASEP, we should obtain the linear profile again.
 Already for $\sigma=3.0$ (Fig. \ref{fig6}) the density profile is almost linear for any system size, 
similarly as in the case of TASEP, and the aim is to find the threshold at which the 
standard TASEP regime sets on.

%
%

\begin{figure}
\includegraphics{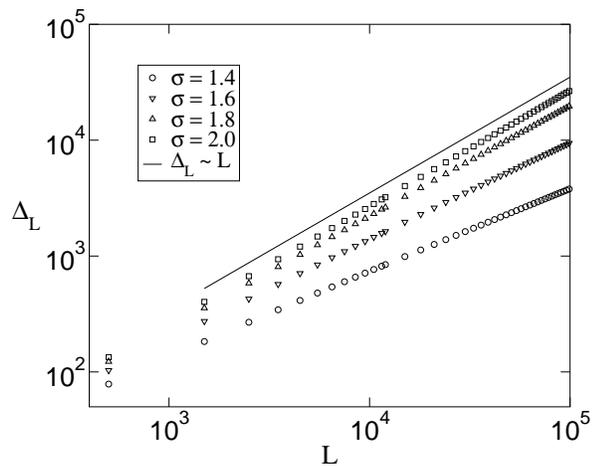}
\caption{Standard deviation of the domain wall position according to the stationary probability
$P_i$ for $\alpha=\beta=0.1$ and various $\sigma$.}
\label{fig7}
\end{figure}

\indent As pointed out in \cite{RakosPaessensSchutz06}, the localization of the random walker is possible 
when the effective potential (\ref{V_i}) is not homogeneous but has minima and maxima. If the potential 
has only one minimum, as in the present model, the localization is possible in the sense that

\begin{equation}\label{stdev}
\lim_{L\rightarrow\infty}\frac{[\langle x^2\rangle_{s}-\langle x\rangle_{s}^2]^{1/2}}{L}=0,
\end{equation}

\noindent where $x$ is the position of the random walker and $\langle\cdots\rangle_{s}$ denotes the average taken over the
stationary probability $P_i$. Using the above definition we plot the standard deviation
$\Delta_L=[\langle x^2\rangle_{s}-\langle x\rangle_{s}^2]^{1/2}$ as a function of the system size $L$
in Fig. \ref{fig7}. We find $\Delta_L$ in good agreement with the power law

\begin{equation}\label{sigma_s}
\Delta_L\sim \left\{\begin{array}{ll}
L^{\frac{\sigma}{2}} & \textrm{$1<\sigma<2$} \\ 
L & \textrm{$\sigma>2$} 
\end{array}\right..
\end{equation}

%
%

\noindent Therefore, the domain wall is localized for $1<\sigma<2$ and delocalized for $\sigma\geq 2$. This result can
be obtained explicitly by taking the continuum limit of the expressions (\ref{p_i}), (\ref{q_i}), (\ref{P_i}), and (\ref{V_i}) and
checking the Taylor expansion of the potential (\ref{V_i}) around the minimum located at the site $i=L/2+1$
(for even system size $L$). Since for $\alpha=\beta$ the symmetry relation $l_i=r_{L-i+2}$ holds, only the even powers
of $(x-1/2)$ are nonvanishing:

\begin{equation}
V(x)=V(1/2)+c_2(x-1/2)^2+c_4(x-1/2)^4+\cdots.
\end{equation}

\noindent It can be shown (see Appendix B) that both $c_2$ and $c_4$ are of the same order in $L$,

\begin{equation}
c_2\sim c_4\sim L^{2-\sigma}.
\end{equation}

\noindent Therefore, if we keep only the quadratic term in the Taylor expansion, we obtain the Gaussian distribution
and  the variance proportional to $1/\sqrt{c_2}$ yields exactly the result (\ref{sigma_s}).

The phase separation has already been encountered in TASEP with ``Langmuir kinetics'', first
for particle creation and annihilation rates $\Omega_{A}$ and $\Omega_{D}$ that depend on the system size as $L^{-1}$
\cite{EvansJuhaszSanten03} and later for $\Omega_{A}, \Omega_{D}\sim L^{-a}$ as long as $1<a<2$
\cite{JuhaszSanten04}. 
In the first case the width of the shock was found to decay as $L^{-1/2}$, whereas in the 
second case the decay was still algebraic but with the exponent $a/2$. 
Since the similarity between these two models and the one we consider here 
lies in the fact that both models have the bulk reservoir, we checked the dependence of 
the rates $\alpha_i$ and $\beta_i$ on the system size $L$ for $\sigma>1$ (see Appendix B). 
In the continuum limit and away from the boundaries, we found a similar 
power-law dependence on the system size, i.e. $\alpha(x),\beta(x)\sim L^{-\sigma}$, 
so that the comparison can be made with TASEP with ``Langmuir kinetics'' and where $a=\sigma$. 
In terms of the characteristic length $\gamma$ (\ref{gamma}), it is clear that the similarity between the present model
and the other two models with bulk reservoirs is established only when $\gamma$ diverges, since in this case the boundaries
have a long-range influence on the bulk dynamics. 

\subsection{Maximum-current phase}

As discussed in Sec. III (Cf.  Fig. \ref{fig4}), the maximum-current phase 
also appears in the presence of long-range hopping when $\sigma > 1$ and sets in by a continuous 
phase transition for the same values of  parameters ($\alpha, \beta$) as in the standard TASEP.
However, one may expect that long-range hopping would alter the long-range correlations and therefore the
behavior at the second-order phase transition to the maximum-current phase.

%
%
\begin{figure}[!ht]
\includegraphics{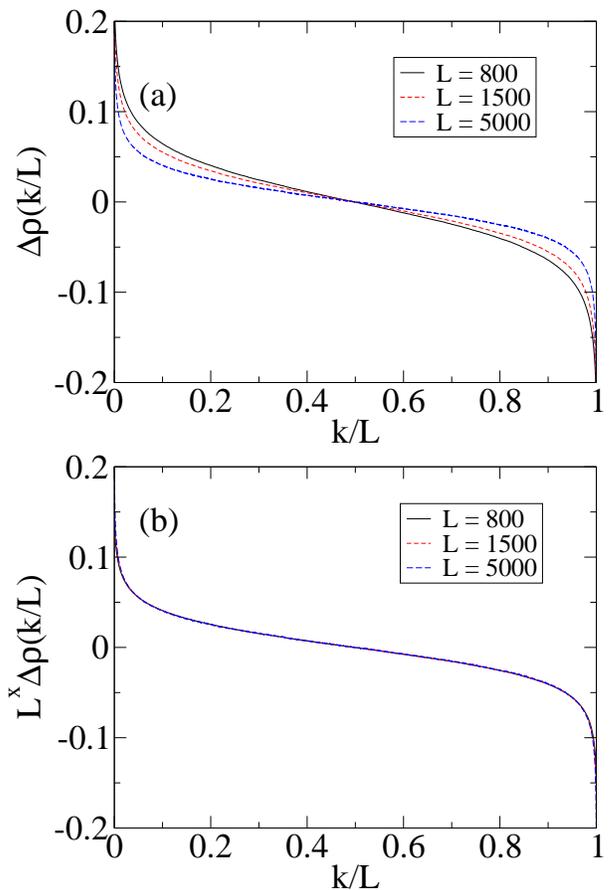}
\caption{(Color online) Deviation $\Delta\rho(k/L)$ of the density profile from its bulk value $\overline{\rho}=1/2$
for $\sigma=1.5$ and various system sizes $L$ ($\alpha=\beta=0.8$), before (a) and after (b) the scaling relation
(\ref{mc_scaling}) was applied.}
\label{fig8}
\end{figure}

The exact solution in the case of TASEP with the nearest-neighbor hopping \cite{SchutzDomany93,DerridaEvans93} shows that 
the transition to the maximum-current phase is characterized by an algebraic decay of the correlation 
function, while the correlation length remains infinite in the entire maximum-current phase.
This is reflected in the algebraic decay of the density profile toward its bulk value $\overline{\rho}=1/2$ with the 
exponent $1/2$.

%
%
\begin{figure}[!ht]
\includegraphics{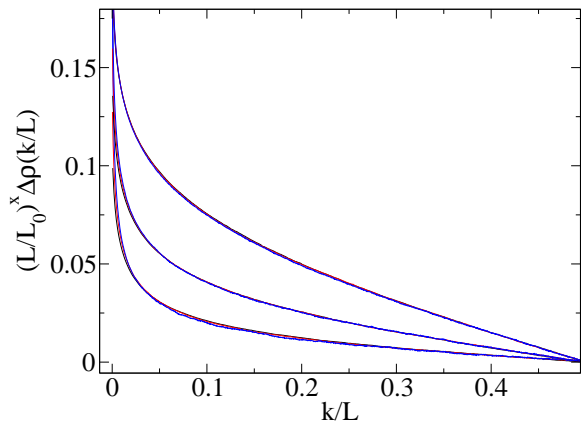}
\caption{(Color online) Deviation $\Delta\rho(k/L)$ of the density profile from its bulk value $\overline{\rho}=1/2$
for various system sizes $L=800$, $1500$, and $5000$ ($\alpha=\beta=0.8$) for $\sigma=1.2$, $1.5$, and $1.8$ 
(from top to bottom). The profiles $\Delta\rho(k/L)$ for the same $\sigma$ are scaled to the 
profile $\Delta\rho(k/L_0)$ ($L_0=5000$) according to Eq. (\ref{mc_scaling}).}
\label{fig9}
\end{figure} 

We studied this decay in the presence of long-range hopping by considering the deviation of a 
profile from its bulk value, $\Delta\rho(k)\equiv\left\vert\langle\tau_k\rangle-1/2\right\vert$,   
by using Monte Carlo simulations. 
We found that for given $\alpha$, $\beta$, and $\sigma$ and for different system sizes $L_1$ and $L_2$, 
$\Delta\rho(k)$ obeys the scaling relation

\begin{equation}
\label{mc_scaling}
\Delta\rho(k/L_1)=\left(\frac{L_2}{L_1}\right)^x\Delta\rho(k/L_2),
\end{equation}

\noindent where $x$ depends continuously on $\sigma$ for $1<\sigma<2$ and is equal 
to the exponent $1/2$ for $\sigma>2$. 
Fig. \ref{fig8}a shows typical results of Monte Carlo simulations ($\sigma=1.5$, $\alpha=\beta=0.8$)
for $\Delta\rho(k/L)$ for various system sizes $L$.
The best numerical fit to the scaling Eq. (\ref{mc_scaling}), for which all data collapse on a single curve (see
Fig. \ref{fig8}b), is achieved for $x=1/4$.
In Fig. \ref{fig9} we present the collapsing fits for several values of $\sigma$. 
The best numerical results for collapsing fits are achieved for $x=(\sigma-1)/2$ which, along 
with $x=1/2$ for $\sigma>2$,  leads to the conjecture

\begin{equation}
x=min\left\{\frac{\sigma-1}{2},\frac{1}{2}\right\},\quad\sigma>1.
\end{equation}

\section{Conclusion}

        We considered the totally asymmetric exclusion process (TASEP) with open boundary conditions,
generalized in such a way that the long-range effect was introduced directly, through long-range hopping,
where the probability of distant hoppings decays as a power law  $1/l^{\sigma+1}$.
Boundary conditions were adjusted in such a way that the periodic boundary conditions are attained
in the special case $\alpha = 1- \beta$.

We found that for $\sigma > 1$ the phase diagram of this generalized model remains the same as the one
in TASEP with the nearest-neighbor hopping, containing the same first- and second-order transition lines.
The bulk density equals the one in the standard TASEP, while the current is equal to the one in TASEP 
normalized by a factor depending on $\sigma$.

As expected, new effects were found at the transition lines, where long range of hopping has 
an impact on long-range correlations which determine the character of the transitions. 
They were observed by studying the density profile.

At the first-order transition line $\alpha = \beta$ the profile becomes localized, in contrast to the
standard TASEP case, but similarly to the model which includes Langmuir kinetics \cite{JuhaszSanten04}.

Like in the standard TASEP, the second-order phase transition to the maximum-current phase has an 
infinite correlation length (as long as $\sigma >1$), which remains infinite in the whole maximum-current 
phase, and the profile displays an algebraic decay. However, the corresponding exponent 
is different from $1/2$ in the region $1<\sigma<2$, where it depends on the range parameter $\sigma$.

In the special case $\sigma = -1$, which belongs to the regime without phase transitions, 
$\sigma < 1$, and corresponds to uniform hopping probability, we derived the exact analytical 
results both for the bulk properties and for the density profile along the line $\alpha = \beta$.

The dynamical scaling, now under consideration, should give a more complete description 
of the considered second-order phase transition. 
It would also be of interest to examine the relation to some other models 
for nonequilibrium phase transitions with long-range interactions, in particular the models 
including temperature such as the two-temperature kinetic Ising model.

\appendix
\section{Case $\sigma=-1$ in the continuum mean-field approximation}

In the continuum limit we set $i/L\rightarrow x$ and substitute all sums in Eq. (\ref{dis_cont_eq}) with integrals.
When $\sigma=-1$, we get simple expressions for the probabilities $\alpha(x)$ and $\beta(x)$:

\begin{equation}\label{ax_bx}
\alpha(x)=\alpha(1-x),\qquad\beta(x)=\beta x.
\end{equation} 

\noindent Inserting (\ref{ax_bx}) in (\ref{dis_cont_eq}) gives the first-order nonlinear differential equation

\begin{equation}\label{rhox_s0}
\frac{dn}{dx}[2n+(\beta-\alpha-1)x+(1-\mu+\alpha)]-n-\alpha(1-x)=0
\end{equation}

\noindent with the boundary conditions 

\begin{gather}
n(0)=0\quad\textrm{and}\quad n(1)=\mu.
\end{gather}

\noindent In the above expressions, $\frac{dn}{dx}=\rho(x)$, and $\mu(\alpha,\beta)=\int_{0}^{1}\rho(x)dx$ is the bulk
density. For $\alpha=\beta$ the first-order differential equation (\ref{rhox_s0}) is \textit{exact}, leading to 

\begin{equation}\label{rhox_s0ab}
\frac{d}{dx}\left\lbrace[n(x)]^2+n(x)\left(\frac{1}{2}+\alpha-x\right)+\left(\frac{1}{2}\alpha x^2-\alpha
x\right)\right\rbrace=0,
\end{equation}

\noindent where we used the fact that $\mu(\alpha,\alpha)=1/2$. 
From (\ref{rhox_s0ab}) the density profile is easily obtained as

\begin{equation}\label{rho_div}
\rho(x)=\frac{1}{2}+\frac{1}{2}\frac{(1-2\alpha)(x-1/2)}{\sqrt{x^2(1-2\alpha)-x(1-2\alpha)+(\alpha+1/2)^2}}.
\end{equation}

\noindent Finally, the total current $j/L$ of particles entering or leaving the chain 
can be obtained by inserting (\ref{rho_div}) into the expression for current, so that

\begin{equation}\label{j_cont}
j/L=\alpha\int_{0}^{1}(1-x)[1-\rho(x)]dx=\beta\int_{0}^{1}x\rho(x),
\end{equation}

\noindent which gives Eq. (\ref{jot_div}).

\section{Asymptotic expansions of $\alpha_i$, $\beta_i$, $r_{i}$ and $l_i$ in the limit of large $L$}

We start by rewriting (\ref{alpha_i}) and (\ref{beta_i}) in a more convenient form:

\begin{equation}
\alpha_i=\alpha\left(1-\frac{\zeta_{i-1}(\sigma+1)}{\zeta_{L}(\sigma+1)}\right),\qquad 
\beta_i=\beta\left(1-\frac{\zeta_{L-i}(\sigma+1)}{\zeta_{L}(\sigma+1)}\right).  
\end{equation}

\noindent Assuming the convergence of the Zeta series, we estimate the partial sum in the limit of large $L$,

\begin{equation}\label{zeta_limit}
\zeta_{L}(\sigma+1)=\zeta(\sigma+1)-\frac{1}{\sigma L^{\sigma}}+o(L^{-\sigma-1}).
\end{equation}

\noindent 
After inserting (\ref{zeta_limit}) in the expressions (\ref{alpha_i}) and (\ref{beta_i}) and 
taking the continuum limit with $x=i/L$, we get for $0\ll x\ll1$ 

\begin{equation}
\alpha(x)\simeq\frac{\alpha}{\zeta(\sigma+1)\sigma}\left[(x-1/L)^{-\sigma}-1\right]L^{-\sigma},
\end{equation}

\begin{equation}
\beta(x)\simeq\frac{\beta}{\zeta(\sigma+1)\sigma}\left[(1-x)^{-\sigma}-1\right]L^{-\sigma}.
\end{equation}

\noindent 
Similarly, the expressions for the hopping rates $r_i$ and $l_i$ in the continuum limit 
for $0\ll x\ll1$ are

\begin{equation}
r(x)\simeq\frac{\beta(1-\beta)}{1-\alpha-\beta}\left[\lambda_{L}+\frac{f(x,\sigma,\alpha,\beta)}{L^{\sigma-1}}\right],
\end{equation}

\begin{equation}
l(x)\simeq\frac{\alpha(1-\alpha)}{1-\alpha-\beta}\left[\lambda_{L}+
\frac{f(1-x+2\epsilon,\sigma,\beta,\alpha)}{L^{\sigma-1}}\right].,
\end{equation}

\noindent 
where $\epsilon=1/L$, and the function $f(x,\sigma,\alpha,\beta)$ is finite for $0<x<1$. 
We define $\Delta_{i}^{(n)}$ as the discrete \textit{n}-th order ``derivation'' of $V_i$  

\begin{equation}
\Delta_{i}^{(n)}=\Delta_{i+1}^{(n-1)}-\Delta_{i}^{(n-1)}
\end{equation}

\noindent with $\Delta_{i}^{(1)}=V_i-V_{i-1}$. In the continuum limit with $x=i/L$ we get 

\begin{equation}
\Delta_{i}^{(n)}\rightarrow L^{-n}\frac{d^n}{dx^n}V(x).
\end{equation}.

\noindent 
Finally, inserting the expressions (\ref{p_i}) and (\ref{q_i}) in the potential (\ref{V_i}) 
and expanding the function $f(x,\sigma,\alpha,\beta)$ around the minimum of the potential, 
$x=1/2+\epsilon/2$, we obtain the result for the first two expansion coefficients:

\begin{equation}
c_2=\frac{1}{2}\left.\frac{d^2}{dx^2}V(x)\right|_{x=1/2}\simeq-\frac{2f'(1/2,\sigma,\alpha,\alpha)}
{\lambda\alpha(1-\alpha)} L^{2-\sigma},
\end{equation}

\begin{equation}
c_4=\frac{1}{24}\left.\frac{d^4}{dx^4}V(x)\right|_{x=1/2}\simeq-\frac{2f'''(1/2,\sigma,\alpha,\alpha)}
{\lambda\alpha(1-\alpha)} L^{2-\sigma}.
\end{equation}

\end{document}